\documentclass[%
 reprint,
 amsmath,amssymb,
 aps,
]{revtex4-2}

\usepackage{graphicx}
\usepackage{bm}
\usepackage{dcolumn}
\usepackage{float}
\usepackage{caption}
\usepackage{subfigure}
\usepackage{amsmath}
\usepackage{mathrsfs}
\usepackage{amssymb}
\usepackage{hyperref}
\usepackage{xcolor}
\let \sss=\scriptscriptstyle

\begin{document}

\preprint{}

\title{Anharmonic interaction as random field for thermal transport in FPU-$\beta$ lattice}
\author{Li Wan}
\email{lwan@wzu.edu.cn}
\affiliation{Department of Physics, Wenzhou University, Wenzhou 325035, P. R. China}%


\begin{abstract}
We present an open quantum theory for the thermal transport in the Fermi-Pasta-Ulam-$\beta$(FPU-$\beta$) lattice. In the theory, local bosons(LBs) are introduced as carriers for the transport. The LBs are stimulated by individual atoms in the lattice, which are different from the phonons that are collective motions of the atoms. The LBs move in the FPU chain and are governed by a set of stochastic differential equations(SDEs). The anharmonic interaction between the atoms in the lattice is transformed to a random field by the Hubbard-Stratonovich transformation, and has been implemented in the set of SDEs. By solving the set of SDEs at the steady state, we study the influence of the anharmonic interaction on the thermal transport. Results show that the anharmonic interaction decreases the thermal current by trapping the LBs on the lattice sites, as well as increase the thermal current by enhancing the amount of the LBs for the transport. The competition between these two mechanisms makes the thermal conductivity of the lattice dependent on the anharmonic interaction non-monotonically. The finite size effect of the thermal conductivity has also been captured by the theory.  
\end{abstract}

\maketitle


\section{introduction}
In open systems, thermal transport in low-dimensional structures is a fundamental topic~\cite{Dhar08,Lepri03}. The thermal transport in the structures breaks the Fourier's Law and is size-dependent, which has been verified by experiments~\cite{Chang08,Chen13,Chen15,Xu14}. Such anomalous behaviors have attracted wide research interests, and have been investigated by various theories~\cite{Dhar08,Lepri03,Weimer21,Simon19,Gardiner00,Ger05,Liu05,Hardy63,Ray19LDT}. However, a complete understanding of the microscopic mechanism for the anomalous transport still remains an issue.\\

In non-metal crystals, lattice vibration is responsible for the thermal transport ~\cite{Born54}. If the interaction between the atoms in the crystals is harmonic, the lattice vibration can be quantized by phonons. The phonons represent collective motions of the atoms in the crystals, and are non-local for any individual atom ~\cite{Born54,Wall72,Ipato71,Sriva90}.  The relation between the frequency and the wave vector of the phonons is the Phonon Dispersion Relation(PDR). In the harmonic crystals, PDRs are lines. The derivative of the frequency with respect to the wave vector along the PDR lines defines the group velocity of the phonons. In the harmonic crystals, the phonons do not interact with each other, and the transport of the phonons is ballistic.\\

In real crystals, the interaction between the atoms is normally anharmonic~\cite{Ziman60,Mara62,Mons13,Liu14,Das14}. The thermal conductivity of the crystals is finite. If the anharmonic interaction is small, the phonons still can be well defined and the anharmonic interaction can be treated by the perturbation method around the harmonic interaction. In this case, the definition of the group velocity of the phonons is still reliable and can be used to study the thermal transport by solving the Phonon Boltzmann Transport Equation(PBTE)~\cite{Ziman60}. The small anharmonic interaction induces  multiphonon scattering, which leads to a finite life time of the phonons and is responsible for the thermal resistance~\cite{Born54,Wall72,Ipato71,Sriva90, Mara62,Ziman60,Liu14}.\\

When the anharmonicity dominates the atomic interactions in the crystals, the PDR lines are broadened, even across each other~\cite{Simon19}.  In these cases, the phonon frequency is not a single-valued function of the wave vector, and the definition for the group velocity of phonons is vague. Additionally, the multiphonon scattering induced by the strong anharmonic interactions makes the computation of the phonon life time very intensive. Thus, the perturbation theory fails and the PBTE is not a proper tool for the study of the thermal transport. A general theory considering the full anharmonic interactions is required. \\

In the phonon space, the open quantum theory(OQT) is a different way to study the thermal transport in the crystals~\cite{Gardiner00,Weimer21,Breuer16, Dhar08}. In the OQT, the crystals are connected to various reservoirs which are set at different temperatures. The temperature drops between the reservoirs drive phonons move from the reservoir with a high temperature to the reservoir with a low one, forming the thermal current. The full Hamiltonian contains not only the crystals but also the reservoirs. In the OQT, the periodical boundary condition for the low dimensional structure is removed, which brings the ambiguity in defining the phonons. Similar problem, the broadening of the PDR lines, also occurs to the OQT if the anharmonic interaction is considered. Additionally, the phonon space will be enormous, and needs a large computational cost. The OQT can be reduced to be stochastic dynamics, such as quantum langevin equation or Fokker-planck equation, if the Markovin approximation is applied~\cite{Wan17,Zeng19}. However, the anharmonic interaction is still a big challenge for the stochastic dynamics.\\

The concept of phonons has also been applied in the molecular dynamics(MD) simulations to investigate the thermal transport in the crystals~\cite{Frenk02,McG06,Sche02,Volz99,Dong01,Turney09,Thomas10,Xu16}. The thermal conductivity can be obtained according to the Green-Kubo(GK) formalism~\cite{Green54,Kubo57,Carbogno17,McG06}. The MD simulations are very flexible because the anharmonicity of the atomic  interactions can be fully considered. However, the MD simulations are classical, and can not reveal the quantum behaviors of the carriers in the thermal transport. Additionally, the MD simulations are reliable for large systems close to the thermodynamic limitation, say the study of the power-law divergence of the thermal conductivity, and are not appropriate to capture the finite size effect of the thermal transport.  \\

In this work, we propose a quantum theory for the study of the thermal transport in a Fermi-Pasta-Ulma-$\beta$(FPU-$\beta$) lattice, which is low-dimensional with anharmonic interactions involved. The theory is in the frame of the OQT, which can mimic the experimental setup. In this theory, we abandon the concept of the phonons as the carriers for the transport. Instead, we define the local boson(LB) on each lattice site. The anharmonic interaction between atoms in the lattice is transformed to a random field by using the field theory. In this way, we needn't define the velocity of the bosons and the enormous boson space is avoided for the calculation. By integrating the Gaussian functions, the influence of the anharmonic interaction on the motion of the LBs in the lattice can be revealed conveniently. Details of the derivation can be found in the supplemental material~\cite{sup}.\\  

\section{theory}
We consider a FPU-$\beta$ lattice having $N$ identical atoms arranged along a straight line periodically. The lattice parameter of the FPU chain is set to 1. The atoms vibrates around their equilibrium positions. For the $j$-th atom, the mass is denoted by $m_j$. The displacement of the atom away from its equilibrium position is denoted by $r_j$, and the momentum of the atom is by $P_j$. The chain contains a system and two reservoirs. The two ends of the system are connected to the two reservoirs respectively. We will specify the system and the reservoirs later, but not at the present stage. The Hamiltonian of the lattice is
\begin{align}
\label{H}
H=\sum_j\frac{P_j^2}{2m_j}+\frac{1}{2}\sum_{(j,k)}B_{jk}r_{jk}^2+\frac{1}{2}\sum_{(j,k)}C_{jk}r_{jk}^4,
\end{align}
with $r_{jk}=r_j-r_k$. The subscript $jk$ of the coefficients $B_{jk}$ and $C_{jk}$ refers to the interaction between the $j$-th and the $k$-th atoms with $j\neq k$. With a pair of given indexes $j$ and $k$, we have two equal quantities $B_{jk}r_{jk}^2$ and $B_{kj}r_{kj}^2$ by swapping $j$ and $k$, which brings the double counting in the sum. To avoid the double counting, we take only one of the two quantities, and use the parentheses $(j,k)$ in Eq.(\ref{H}) to indicate the avoidance. Such treatment has also been applied to the anharmonic term $C_{jk}r_{jk}^4$ in Eq.(\ref{H}). We keep the factor $1/2$ in the anharmonic term for  easy transformation later. For clarity, we denote the terms related to the harmonic interaction in the Hamiltonian by $H_0=\sum_j P_j^2/(2m_j)+(1/2)\sum_{(j,k)}B_{jk}r_{jk}^2$.\\

\subsection{Path Integral Method}
By using the path-integral method, we obtain the wave function $|\Psi_l>$ of the system at the $l$-th time slice evolving from the wave function $|\Psi_{l-1}>$ at the $(l-1)$-th time slice, through the evolving equation $|\Psi_{l}>=\mathcal{P}_l|\Psi_{l-1}>$. The path-integral propagator is
\begin{align}
\label{PI}
\mathcal{P}_l=e^{-i\tau [H_0]_l /\hbar}\left[\prod_{(j,k)}e^{-i\frac{\tau}{2}[C_{jk}r_{jk}^4] /\hbar}\right]_l,
\end{align} 
with $\tau$ the time interval. The subscript $l$ means the  $l$-th time slice for the propagator in the path. The quartic term $r_{jk}^4$ in Eq.(\ref{PI}) can be reduced to a quadratic term by the Hubbard-Stratonovich(HS) transformation, showing explicitly as
\begin{align}
e^{-i\frac{\tau}{2}[C_{jk}r_{jk}^4] /\hbar}=\frac{1}{Z_{jk}}\int \left[\mathcal{D}\beta_{\sss jk}\right]e^{-\frac{\tau\beta_{\sss jk}^2}{2}} e^{-i \tau \epsilon c_{\sss jk}'\beta_{\sss jk}r_{jk}^2/\hbar}.
\end{align}
In the above transformation, $i$ is the imaginary unit. $\beta_{\sss jk}$ is the random field introduced by the HS transformation. $[\mathcal{D}\beta_{\sss jk}]$ is the measure of the random field. $\epsilon=\frac{\sqrt{2}}{2}(1+i)$ originates from $\sqrt{i}$, and its conjugate value is $\epsilon^*=\frac{\sqrt{2}}{2}(1-i)$. We denote $c_{\sss jk}'=\sqrt{C_{jk}\hbar}$. The random field $\beta_{\sss jk}\sqrt{\tau}$ follows the Gaussian distribution $\mathcal{N}(0,1)$ with the mean zero and the variance unity. $Z_{jk}=\int \left[\mathcal{D}\beta_{_{jk}}\right]e^{-\tau \beta_{\sss jk}^2/2}$ is the normalization factor. Thus, the propagator expressed in terms of the random fields reads
\begin{align}
\mathcal{P}_l= \int\left[\prod_{(j,k)}\frac{1}{Z_{jk}}[\mathcal{D}\beta_{jk}]e^{-\frac{\tau\beta_{jk}^2}{2}}\right]_l\times e^{-i\tau\mathcal{H}_l/\hbar},
\end{align}
with
\begin{align}
\mathcal{H}=\sum_j\frac{P_j^2}{2m_j}+\frac{1}{2}\sum_{(j,k)}B_{jk}r_{jk}^2+\epsilon\sum_{(j,k)} c_{jk}'\beta_{jk}r_{jk}^2.
\end{align}
We emphasize that $\mathcal{H}$ is not the Hamiltonian of the system. It is non-hermitian and random. For simplicity, we denote $H_1=\epsilon\sum_{(j,k)} c_{\sss jk}'\beta_{\sss jk}r_{\sss jk}^2$. Thus, we have $\mathcal{H}=H_0+H_1$.\\

\subsection{Local Bosons}
We express $H_0$ in terms of $r_j$ by using $r_{jk}=r_j-r_k$, and get
\begin{align}
H_0=\sum_j\frac{P_j^2}{2m_j}+\frac{1}{2}\sum_{j}B_{j}r_j^2-\frac{1}{2}\sum_{j,k}B_{jk}r_jr_k.
\end{align}
In the above derivation, the restriction $(j,k)$ to avoid the double counting has been removed, which will benefit the functional integration later. $B_j=\sum_{k}B_{jk}$ has also been defined with $k\neq j$. We treat the anharmonic term $H_1$ similarly and obtain
\begin{align}
H_1=\sum_{j}(\sum_k \epsilon c_{jk}'\beta_{jk})r_j^2-\sum_{j,k} \epsilon c_{jk}'\beta_{jk}r_jr_k.
\end{align}
We introduce bosons localized on each lattice site of the chain for the thermal transport. The bosons referred to as local bosons(LBs) are stimulated by the atoms vibrating around their own equilibrium positions individually. LBs are different from Phonons. The latter are global and collective motions of the atoms of the chain. The frequency of the LBs at the $j$-th lattice site is defined by $\omega_j=\sqrt{B_j/m_j}$. The creation and the annihilation operators of the LBs at the $j$-th lattice site read
\begin{align}
\label{aoperator}
a_j^{\dagger}=\sqrt{\frac{\omega_j m_j}{2\hbar }}( r_j-i\frac{P_j}{\omega_j m_j}),~~a_j=\sqrt{\frac{\omega_j m_j}{2\hbar }}( r_j+i\frac{P_j}{\omega_j m_j}).
\end{align}
From the above equations, $r_j$ and $P_j$ can be expressed in terms of $a_j$ and $a_j^{\dagger}$, and then are substituted into $\mathcal{H}=H_0+H_1$. After the rotational approximation, we obtain
\begin{align}
\label{matH}
\mathcal{H}=&\hbar\sum_j  (\omega_j+2\epsilon F_j) a^{\dagger}_ja_j
+\hbar\sum_{jk}2(U_{jk}+\epsilon G_{jk})a_j^{\dagger}a_k\nonumber\\
&+\hbar\epsilon\sum_{j}F_j+\frac{\hbar}{2}\sum_j \omega_j
\end{align}  
with
\begin{align}
\label{UFG}
&U_{jk}=-\frac{B_{jk}}{4\sqrt{\omega_j m_j\omega_km_k}},\nonumber\\
&G_{jk}=-\frac{ c_{jk}'\beta_{jk}}{2\sqrt{m_j\omega_jm_k\omega_k}}.\nonumber\\
&F_{j}=\frac{\sum_k  c_{jk}'\beta_{jk}}{2\sqrt{m_j\omega_jm_j\omega_j}}.
\end{align}
The last term $\frac{\hbar}{2}\sum_j \omega_j$ in Eq.(\ref{matH}) is a constant, and will be omitted in the following.\\

\subsection{Coherent State Path Integral}
We define the coherent state of the FPU-$\beta$ chain by
\begin{align}
|\Xi>=e^{\sum_j \xi_j a^{\dagger}_j}|0>=\prod_j[\sum_{n_j=0}^{\infty}\frac{\xi_j^{n_j}}{\sqrt{n_j!}}]|n_j>
\end{align}
with $n_j$ representing the number of LBs at the $j$-th lattice site. The eigen equations are $a_j|\Xi>=\xi_j|\Xi>$ and $<\Xi|a_j^{\dagger}=<\Xi|\xi_j^*$. According to the coherent state path integral formalism, the unit operator is $\int [\prod_j\mathcal{D}\xi_j\cdot \mathcal{D}\xi_j^* ]e^{-\sum_j\xi_j^*\xi_j}|\Xi><\Xi|=\hat{1}$. We apply the unit operator to represent the propagator $\mathcal{P}_l$ in the LB space by $
\mathcal{P}_l=\hat{1}_l\mathcal{P}_l\hat{1}_{l-1}$. We rewrite the propagator $\mathcal{P}_l$ in the LB space as
\begin{align}
\mathcal{P}_l&=\int [\prod_j\mathcal{D}\xi_{j}^*]_l~[\prod_j\mathcal{D}\xi_j]_{l-1} \nonumber\\
&\times\left[\prod_{(j,k)}\frac{1}{Z_{jk}}[\mathcal{D}\beta_{jk}]e^{-\frac{\tau\beta_{jk}^2}{2}}\right]_l\left[e^{S_l}\right]~
\end{align}
with
\begin{align}
e^{S_l}=e^{-\sum_j\xi_{j,l}^*\xi_{j,l}}<\Xi_l|e^{-i\tau\mathcal{H}_l/\hbar}|\Xi_{l-1}>.\nonumber
\end{align}
The action $S_l$ can be obtained analytically, which can be found in the supplemental material~\cite{sup}. The functional integration of $[\prod_j\mathcal{D}\xi_{j}^*]_l$ in $\mathcal{P}_l$ gives a delta functional, which leads to a stochastic differential equation reading
\begin{align}
\label{dotxi}
\dot{\xi_{j}}=-i \sum_k(\omega_j \delta_{j,k}+2U_{jk})\xi_{k}-i\epsilon \sum_k c_{jk} \beta_{jk} (\xi_{j}-m_{jk}\xi_{k}).
\end{align}
Here, $\delta_{j,k}$ is the Kronecker function. In the above equation, we have defined $c_{jk}=c_{jk}'/(m_j\omega_j)$ and $m_{jk}=\sqrt{m_j\omega_j/m_k\omega_k}$. In the atomic chain, $\beta_{jk}$ is the random field induced by the interaction between the $j$-th atom and the $k$-th atom, and $\beta_{jj}=0$ is always held since there is no self-interaction. \\

To demonstrate our theory, we only consider the interactions between the nearest neighbor atoms. Therefore, we have $c_{j(j+1)}=c_{j(j-1)}=c$ and $m_{jk}=1$, and simplify Eq.(\ref{dotxi}) as
\begin{align}
\label{xi}
\dot{\xi_{j}} 
=-i \sum_k(\omega_j \delta_{j,k}+2U_{jk})\xi_{k}-i\epsilon c\sum_k \beta_{jk} (\xi_{j}-\xi_{k}).
\end{align}
For the conjugate part, we have an equation
\begin{align}
\label{gammastar}
\dot{\gamma_{p}^*} 
=i \sum_q(\omega_p \delta_{p,q}+2U_{pq})\gamma^*_{q}+i\epsilon^* c\sum_q \alpha_{pq} (\gamma^*_{p}-\gamma^*_{q}).
\end{align}
Here, we use the notions $\gamma^*$ and $\alpha$ to differ from $\xi$ and $\beta$ respectively for clarity. The two stochastic differential equations Eq.(\ref{xi}) and Eq.(\ref{gammastar}) are the basis for the study of the LBs moving in the chain.\\

\subsection{Quantum Average}
By using Eq.(\ref{xi}), the propagator $\mathcal{P}_l$ is reduced to be a simple form. Suppose the coherent state of the chain is initialized at
$|\Xi(0)>$. After a duration time $\tau$, the coherent state of the chain evolves to a new state by $|\Xi(\tau)>=\mathcal{P}_{l}|\Xi(0)>$. Due to the over-completeness properties of the coherent states, we have
\begin{align}
&<\Xi(\tau)|\Xi(\tau)>=<W>\nonumber\\
&=<e^{ic\tau [\sum_{(p,q)}\epsilon^* \alpha_{pq}-\sum_{(j,k)}\epsilon\beta_{jk}]}e^{\sum_{j}\gamma_j^*(0)\xi_j(0)}>.
\end{align}
Here, we use $W$ for the simple notation, and the quantum average $<W>$ is referred to as
\begin{align}
\label{W}
&<W>=\int~~ W\nonumber \times \nonumber\\
&\left[\prod_{(p,q)}\frac{1}{Z_{pq}}[\mathcal{D}\alpha_{pq}]e^{-\frac{\alpha_{pq}^2}{2}\tau}\right]\left[\prod_{(j,k)}\frac{1}{Z_{jk}}[\mathcal{D}\beta_{jk}]e^{-\frac{\beta_{jk}^2}{2}\tau}\right].
\end{align}
For a given operator $\hat{O}$, the quantum average of the operator $\hat{O}$ is
\begin{align}
\label{qavg}
\overline{O}=\frac{<\Xi(\tau)|\hat{O}|\Xi(\tau)>}{<\Xi(\tau)|\Xi(\tau)>}=\frac{<O(\alpha,\beta)W>}{<W>}.
\end{align}
The quantum averages of the operators can be realized by the Monte Carlo method, and the random fields $\beta$ and $\alpha$ are the Gaussian random numbers according to Eq.(\ref{W}). Fortunately, the quantum averages of the correlation functions can be analytically obtained. The quantum average of the operator $a_j^{\dagger}a_j$ means the LB number at the $j$-th lattice site, which can be obtained by $\overline{\gamma_j^*\xi_j}=<a_j^{\dagger}a_jW>/<W>$. And $\overline{\gamma_j^*\xi_k}=<a_j^{\dagger}a_kW>/<W>$ means the rate of the LBs hopping from the $k$-th lattice site to the $j$-th lattice site.\\

\subsection{Transport Equation of LBs}
In the stochastic process, $\dot{\xi}$ is represented by $\Delta{\xi}/\tau$ and $\dot{\gamma^*}$ by $\Delta{\gamma^*}/\tau$. The increments $\Delta{\xi}$ and $\Delta{\gamma^*}$ can be obtained from Eq.(\ref{xi})and Eq.(\ref{gammastar}). According to the theory of the stochastic process, we have $\Delta (\gamma_p^*\xi_j)=(\Delta \gamma_p^*)\xi_j+\gamma_p^*(\Delta \xi_j)+(\Delta \gamma_p^*)(\Delta \xi_j)$, and then get a differential equation for $\gamma_p^*\xi_j$ through $\dot{(\gamma_p^*\xi_j)}=\Delta(\gamma_p^*\xi_j)/\tau$. We go further to make the quantum average over $\dot{(\gamma_p^*\xi_j)}$, which leads to 
\begin{align}
\label{avgequ}
\overline{\dot{\gamma_{p}^*\xi_j}}&=\overline{\frac{\Delta(\gamma_{p}^*\xi_j)}{\tau}}=\overline{B_{pj}^1}~~\gamma_p^*\xi_j+\sum_q\overline{B_{qj}^2}~~\gamma_q^*\xi_j\nonumber\\
&+\sum_k\overline{B_{pk}^3}~~\gamma_p^*\xi_k+\sum_{q,k}\overline{B_{qk}^4}~~\gamma_q^*\xi_k,
\end{align}
with the coefficients of 
\begin{align}
\label{avgequB}
&\overline{B_{pj}^1}=c^2\tau\sum_{q,k}\overline{\alpha_{pq}\beta_{jk}},\nonumber\\
&\overline{B_{qj}^2}=iA_{pq}-ic^2(1-\delta_{pq})+2c^2\tau A_{pq}-c^2\tau\sum_k \overline{\alpha_{pq}\beta_{jk}},\nonumber\\
&\overline{B_{pk}^3}=-iA_{jk}+ic^2(1-\delta_{jk})+2c^2\tau A_{jk}-c^2\tau\sum_q \overline{\alpha_{pq}\beta_{jk}},\nonumber\\
&\overline{B_{qk}^4}=\tau A_{pq}A_{jk}-c^2\tau [A_{pq}(1-\delta_{jk})+A_{jk}(1-\delta_{pq})]\nonumber\\
&~~~~~~~~+c^2\tau\overline{\alpha_{pq}\beta_{jk}}.
\end{align}
In the above definitions, we use the superscripts to distinguish the coefficients $B$ in Eq.(\ref{avgequ}) from the coefficients $B$ in Eq.(\ref{H}) for saving symbols. The quantities $\gamma^*\xi$ on the right hand side of Eq.(\ref{avgequ}) are interpreted in ito sense, and have already been quantum-averaged before the duration time $\tau$.\\

At the steady state, $\overline{\gamma_p^*\xi_j}$ is time independent and $\overline{\Delta(\gamma_p^*\xi_j)}$ equals zero. Therefore, Eq.(\ref{avgequ}) can be viewed as a set of linear equations. And all the variables $\overline{\gamma_j^*\xi_k}$ can be solved out if appropriate conditions are imposed.\\

\subsection{Continuity Equations}
We focus on the LB number on the $j$-th lattice site, and set $p=j$ in Eq.(\ref{avgequ}). In this way, the quantity $\overline{\dot{\gamma_{j}^*\xi_j}}$ on the left hand side of Eq.(\ref{avgequ}) is the changing rate of the LB number while the right hand side of Eq.(\ref{avgequ}) represents the thermal currents flowing into the $j$-th lattice site as well as leaving the site. Thus, Eq.(\ref{avgequ}) is the continuity equation for the LBs on the $j$-th lattice site after the setting of $p=j$. To show the thermal currents clearly, we rearrange the terms of Eq.(\ref{avgequ}) in a different form, reading
\begin{align}
\label{Jinout}
\overline{\dot{\gamma_{j}^*\xi_j}}=J_{(<j)}+J_{(=j)}+J_{(>j)}.
\end{align}
Here, $J_{(<j)}$ is the thermal current flowing from the atoms with the indexes smaller than $j$ to the $j$-th atom, reading
\begin{align}
\label{Jin}
J_{(<j)}=D_{j11}~\gamma_{j-1}^*\xi_j+D_{j12}~\gamma_{j}^*\xi_{j-1}+D_{j13}~(\gamma_{j-1}^*\xi_{j-1}-\gamma_{j}^*\xi_{j}),
\end{align}
with $D$ the coefficients derived from the coefficients $B$ of Eq.(\ref{avgequB}).
In the above definition, the first and the second terms represent the LBs hopping between the $j$-th and the $(j-1)$-th lattice sites. The third term is responsible for the LBs diffusing in the lattice, which is dependent on the gradient of the LB numbers between the two sites. The term $J_{(>j)}$ in Eq.(\ref{Jinout}) is the thermal current flowing from the atoms with larger indexes than $j$ to the $j$-th atom, which reads
\begin{align}
\label{Jout}
J_{(>j)}=D_{j21}~\gamma_{j+1}^*\xi_j+D_{j22}~\gamma_{j}^*\xi_{j+1}+D_{j23}~(\gamma_{j+1}^*\xi_{j+1}-\gamma_{j}^*\xi_{j}).
\end{align}
Similarly, the first and the second terms represent the LBs hopping between the $j$-th and the $(j+1)$-th lattice sites. The third term is for the thermal current by the diffusion of the LBs in the lattice. The currents $J_{(<j)}$ and $J_{(>j)}$ are negative to each other. The quantity $J_{(<j)}+J_{(>j)}$ is the net thermal current flowing to the $j$-th atom from all the other atoms in the chain. In this study, we have set the lattice parameter of the atomic chain as unit. Generally, the thermal current is $J_{(<j)}$ (or $J_{(>j)}$ ) multiplying the lattice parameter.\\

In Eq.(\ref{Jinout}), $J_{(=j)}$ is the thermal current for the LBs created or annihilated on the $j$-th lattice site itself, which is 
\begin{align}
\label{Jequj}
J_{(=j)}=D_{j31}~\gamma_{j}^*\xi_j+D_{j32}\gamma_{j+1}^*\xi_{j-1}+D_{j33}~\gamma_{j-1}^*\xi_{j+1}.
\end{align}
The last two terms in Eq.(\ref{Jequj}) originate from the operators $a_{j+1}^{\dagger}a_{j-1}$ and $a_{j-1}^{\dagger}a_{j+1}$ respectively, showing that the LBs hop between the $(j-1)$-th and the $(j+1)$-th lattice sites by passing through the $j$-th atom, which creates or annihilates the LBs at the $j$-th lattice sites. The coefficients $D$ in Eq.(\ref{Jin}), Eq.(\ref{Jout}) and Eq.(\ref{Jequj})can be found in the supplemental material~\cite{sup}.\\

At the steady state, the LB number should be conserved in the whole chain, and no extra LBs are created or annihilated at each lattice site. The conservation of the LB number in the chain requires that the number of LBs hopping to the $j$-th atom equals the number of LBs leaving the atom, expressed by $J_{(<j)}+J_{(>j)}=0$. Equivalently, it is
\begin{align}
\label{contiequ21}
J_{(=j)}=0
\end{align}
at the steady state on each lattice site. Similarly, the number of LBs leaving the $j$-th atom equals the number of LBs flowing to its neighbor atom, which requires  
\begin{align}
\label{contiequ22}
J_{(>j)}+J_{(<(j+1))}=0.
\end{align}
Now we have one set of equations Eq.(\ref{avgequ}) with $\overline{\dot{\gamma_p^*\xi_j}}=0$ at the steady state, and two requirements  Eq.(\ref{contiequ21}) and Eq.(\ref{contiequ22}). \\

\subsection{Correlations of Random Fields}
\label{randfield}
If the random fields $\alpha_{pq}$ and $\beta_{jk}$ are independent of each other, we get the quantum average of the correlation $\overline{\alpha_{pq}\beta_{jk}}=\overline{\alpha_{pq}}~\times~\overline{\beta_{jk}}=c^2(1-\delta_{pq})(1-\delta_{jk})$ directly~\cite{sup}. However, such result does not satisfy the requirement Eq.(\ref{contiequ21}). Thus, the dependence between the random fields must be considered. For this, we suggest a joint probability density function(JPDF) of two random fields $x$ and $y$ by 
\begin{align}
\label{jpd}
P(x,y)\sim e^{-\frac{\tau}{2(1-\rho_{x,y}^2)}[x^2+y^2-2\rho_{x,y} xy]}
\end{align}
with $\rho_{x,y}$ the correlation coefficient(CC). $P(\alpha_{j(j-1)},\beta_{j(j+1)})$ and $P(\alpha_{j(j+1)},\beta_{j(j-1)})$ share the same CC that is denoted by $\rho_j$ for the $j$-th lattice site. The conditional probability density function is obtained as
\begin{align}
\label{cpd}
P(y|x)\sim e^{-\frac{\tau}{2(1-\rho_{x,y}^2)}[y-\rho_{x,y} x]^2},
\end{align}
from Eq.(\ref{jpd}).\\

According to the Chapman-Kolmogorov Equation(CKE), we can calculate CCs for any two given random fields. The result indicates that the CC of two given random fields is the product of all the CCs of the random fields connecting the two given ones. Explicitly, by using the CKE  $P(c|a)=\int P(c|b)P(b|a)db$, we have $\rho_{\sss c,a}=\rho_{\sss c,b}\rho_{\sss b,a}$. The correlation between two random fields is obtained by $\overline{\alpha_{jk}\beta_{pq}}=<W\cdot\alpha_{jk}\cdot\beta_{pq}\cdot P(\alpha_{jk},\beta_{pq})>/<W\cdot P(\alpha_{jk},\beta_{pq})>$. On the basis of this result, we obtain that the CC for $P(\alpha_{j(j-1)},\beta_{j(j-1)})$ equals $-1$~\cite{sup}.\\

By integrating Gaussian functions, we obtain some analytical results, such as
\begin{align}
&\overline{\alpha_{j(j-1)}\beta_{j(j+1)}}=\overline{\beta_{j(j-1)}\alpha_{j(j+1)}}=\frac{\rho_j}{\tau}+c^2(1+\rho_j^2),\nonumber\\
&\overline{\alpha_{j(j-1)}\beta_{j(j-1)}}=\overline{\alpha_{j(j+1)}\beta_{j(j+1)}}=-\frac{1}{\tau}+2c^2.
\end{align}
All the other correlations of two random fields can be worked out according to the above analytical results. We emphasize that the CC $\rho_j$ must be in the range of $[-1,1]$ in the numerical study. \\

\subsection{Algorithm}
The FPU chain contains a system and two reservoirs. The two ends of the system are connected to the two reservoirs respectively. The two reservoirs are set at two different temperatures to form a temperature drop. At the steady state, we have $\overline{\dot{\gamma_p^* \xi_j}}=0$, which simplifies  Eq.(\ref{avgequ}) to be
\begin{align}
\label{linearequ}
&\overline{B_{pj}^1}~~\gamma_p^*\xi_j+\sum_q\overline{B_{qj}^2}~~\gamma_q^*\xi_j+\sum_k\overline{B_{pk}^3}~~\gamma_p^*\xi_k\nonumber\\
&+\sum_{q,k}\overline{B_{qk}^4}~~\gamma_q^*\xi_k=0.
\end{align}
The number of LBs at each lattice site in the reservoirs is fixed by the Bose-Einstein distribution, which is denoted by $\gamma^*_r\xi_r$. The LB numbers $\gamma^*_r\xi_r$ in the reservoirs need to be moved to the right hand side of Eq.(\ref{linearequ}), forming a set of linear equations. To study the thermal transport in the system of the chain at the steady state, we need to solve Eq.(\ref{linearequ}) imposed by two conditions Eq.(\ref{contiequ21}) and Eq.(\ref{contiequ22}). We emphasize that the equations $\overline{\dot{\gamma_r^* \xi_r}}=0$ should be dropped off for the set of the equations, because the thermal state of the reservoirs are controlled externally. The algorithm is listed in the following.\\

\begin{itemize}
\item{Step1. Set a value for the duration time $\tau$. Set all the correlation coefficients $\rho_j=0$. Solve Eq.(\ref{linearequ}) to get $\gamma_p^*\xi_j$ for the whole system. By using the results of $\gamma_p^*\xi_j$, solve Eq.(\ref{contiequ21}) to get $\rho_j$. And then use the results of $\rho_j$ to solve Eq.(\ref{linearequ}) again to update $\rho_j$ iteratively. Repeat the iteration and record the converged results. In this way, the requirement Eq.(\ref{contiequ21}) is satisfied.}
\item{Step2. Calculate $J_{(<j)}$ for the thermal current by using Eq.(\ref{Jin})(or $J_{(>j)}$ alternatively by Eq.(\ref{Jout})). Average the currents $\mathcal{J}=\sum_jJ_{(<j)}/N$, which is calculated only for the system with the reservoirs excluded. Calculate the deviation of the currents by $\Delta \mathcal{J}=\sum_jJ_{(<j)}^2/N-\mathcal{J}^2$, which is still calculated for the system without the reservoirs.}
\item{Step3. Vary $\tau$ and repeat Step 1 and Step 2 in order to find out the minimum $\Delta \mathcal{J}$. In this way, the requirement Eq.(\ref{contiequ22}) is satisfied. There may exist several minimums $\Delta \mathcal{J}$ for various $\tau$. Choose the smallest $\tau$.}
\end{itemize}
After the steps mentioned above, we can collect data of the thermal currents and study the properties of the thermal transport.\\

\section{results}
For the numerical study, we set the mass of each atom and the lattice parameter of the FPU chain both to be unit. The coefficients $B_{jk}$ in the Hamiltonian Eq.(\ref{H}) are fixed to be $1$. The frequency of the LBs is then calculated to be $\omega=1.414$. The anharmonic coefficients $C_{jk}$ in Eq.(\ref{H}) are also considered only for the nearest neighbor atoms, and are represented by $C$ for simple notation. In this study, $C$ varies from $0.1$ to $0.8$. The FPU chain comprises a system and two reservoirs. The system has $L$ atoms. The first atom of the system is indexed to be $1$, which is connected to the reservoir with a high temperature $T_H$. The low temperature of the reservoir connected to the $L$-th atom of the system is denoted by $T_L$. According to the Bose-Einstein distribution, the LB number at each lattice site of the reservoirs is obtained through $n=1/(e^{\hbar\omega/(k_BT)}-1)$. We use $\hbar \omega_0=k_B T_0=1kJ/mol$ for the normalizations of the frequency $\omega_0$ and the temperature $T_0$. In this study, we fix $T_H=3.8$ and $T_L=2.3$. Thus, the LB numbers are obtained to be $n_H=2.22$ and $n_L=1.18$ for the reservoirs respectively. \\

\subsection{Distribution of LB Numbers}
\begin{figure}[!h]
\includegraphics[scale=0.8]{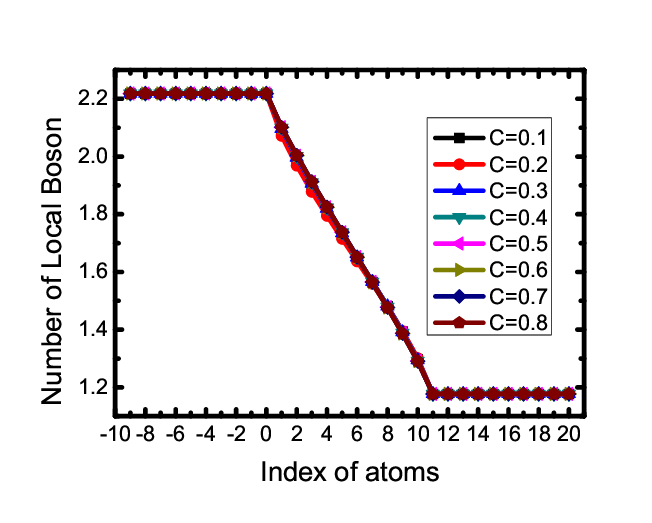}
\caption{Linearized distribution of the LB numbers in the system of $L=10$. The atoms of the system are indexed from 1 to 10. The atoms of the reservoirs have also been indicated. The data for various C overlap.}
\label{fig1}
\end{figure}

Since the temperature can not be defined in the non-equilibrium system, we use the LB numbers along the system to indicate the non-equilibrium state. It is obtained that the LB numbers are linearized in the system. We take the results of $L=10$ as an example, and plot the results in Fig.~\ref{fig1}.
It is found in the figure that the data for various $C$ overlap, showing that the distribution of the LB numbers in the system is determined by the temperature drop of the reservoirs and independent of the anharmonic coefficient $C$ at the steady state. \\

\subsection{$\tau$ and $\rho$}
\label{taurho}
The steady state of the system is a macro-phenomenon. Microscopically, the thermal current is in fluctuation. The duration time $\tau$ is the time scale for the fluctuation. The term $\overline{\dot{\gamma_j^*\xi_j}}=\Delta (\gamma_j^{*}\xi_j)/\tau$ in Eq.(\ref{Jinout}) means the changing rate of the LB number on the $j$-th lattice site. For the steady state, the LBs accumulate on the lattice sites and then leave the sites in the duration time $\tau$. Therefore, $\tau$ is the duration for the LBs staying on the lattice sites. In Eq.(\ref{avgequ}), the parameter $\tau$ is kept in the coefficients shown in Eq.(\ref{avgequB}), and is not omitted by taking its limitation of zero since $\tau$ has the clear physical meaning for the thermal fluctuations.\\

\begin{figure}[!h]
\includegraphics[scale=0.8]{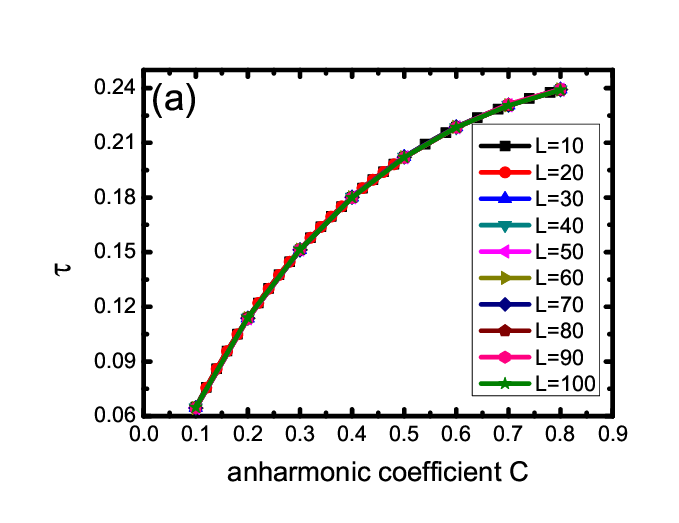}
\includegraphics[scale=0.8]{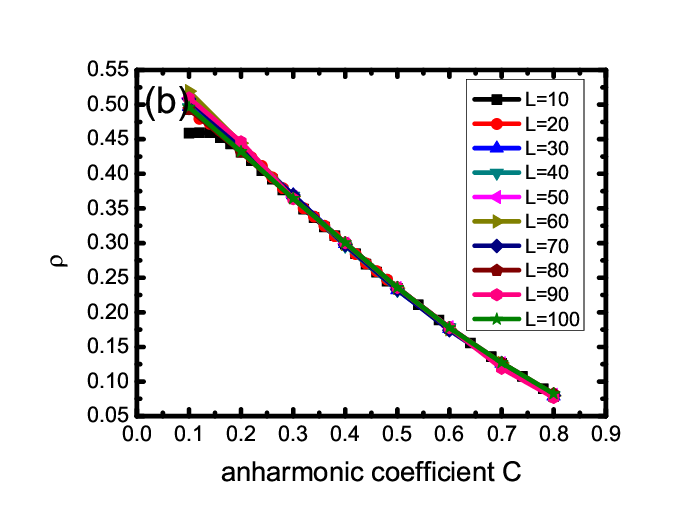}
\caption{Influence of the anharmonic coefficient $C$ on the transport of the LBs. (a) The anharmonic interaction blocks the motion of the LBs in the system and tends to trap the LBs on the lattice sites. (b) The anharmonic interaction weakens the correlations between the atoms.}
\label{fig2}
\end{figure}

In Fig.~\ref{fig2}(a), we plot $\tau$ as a function of the anharmonic coefficient $C$ for various lengths of the system. 
It indicates in the figure that all the data for the various lengths $L$ of the system are overlapping, showing that $\tau$ is independent of the length $L$ of the system. It is an intrinsic parameter of the FPU chain determined by the anharmonic coefficient $C$. With the increasing of $C$, $\tau$ is increased, meaning that the anharmonic interaction blocks the moving of the LBs and traps the LBs on the lattice sites for the duration time $\tau$. It is excepted that the thermal current is decreased with the increase of $\tau$.\\

We have introduced the correlation coefficient $\rho_j$ for the $j$-th lattice site in Sec.(\ref{randfield}). We average the correlation coefficients $\rho_j$ over the whole system and use $\rho$ to denote the averaged value. In Fig.~\ref{fig2}(b), we plot $\rho$ as a function of $C$. Similar to the results in Fig.~\ref{fig2}(a), $\rho$ is also an intrinsic parameter of the system dependent on $C$, and is independent of the length $L$ of the system. With the increase of $C$, $\rho$ is decreased, meaning that less correlation occurs to the atoms in the system with a larger value of the anharmonic coefficient $C$.\\

For the harmonic lattice with $C=0$, the lattice vibration is collective. The collective motion of the atoms in the system correlates the atoms with the largest $\rho$ shown in Fig.~\ref{fig2}(b), and helps the LBs to transport in the system. In this harmonic case of $C=0$, the duration time $\tau$ for the trapping of the LBs on the lattice sites is the smallest, as shown in Fig.~\ref{fig2}(a). The increase of the anharmonic coefficient $C$ weakens the correlation between the atoms. In this way, $\rho$ is decreased, and $\tau$ is increased, which has been revealed in Fig.~\ref{fig2}. 

\subsection{Thermal Conductivity}
\begin{figure}[!h]
\includegraphics[scale=0.8]{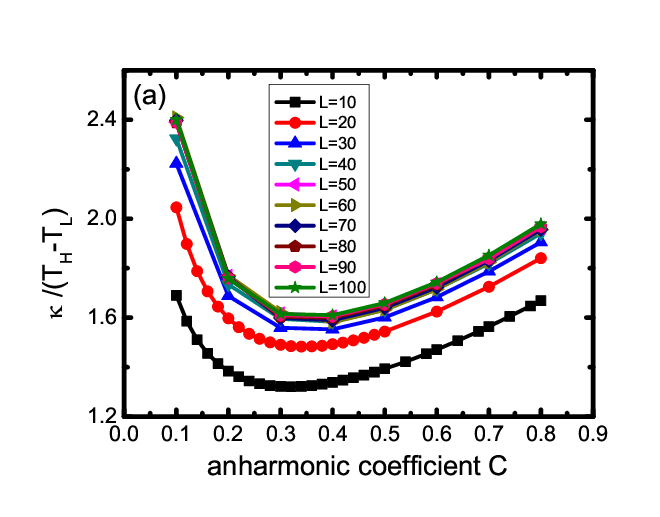}
\includegraphics[scale=0.8]{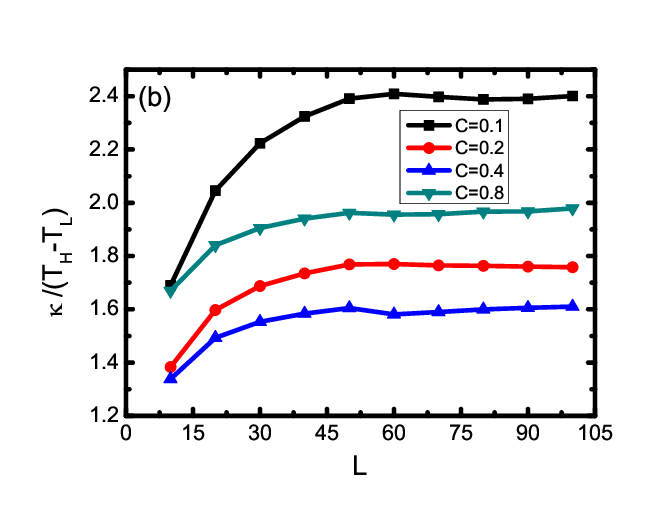}
\caption{Thermal conductivity $\kappa$ as functions of the anharmonic coefficient $C$ and the length $L$ of the system. (a)$\kappa$ is dependent on the anharmonic coefficient $C$ non-monotonically. (b)The finite size effect of $\kappa$ has been observed.}
\label{fig3}
\end{figure}

According to the Fourier's Law, the thermal conductivity $\kappa$ can be obtained by $\kappa=|J_{(<j)}|L/(T_H-T_L)$ in this study. We plot the results of $\kappa$ in Fig.~\ref{fig3} by taking $1/(T_H-T_L)$ as unit.   
Fig.~\ref{fig3}(a) plots $\kappa$ as a function of the anharmonic coefficient $C$. It is observed in the figure that $\kappa$ decreases first and then increases with the increase of $C$. We have revealed in Fig.~\ref{fig2} that the increase of the anharmonic interaction tends to trap the LBs on the lattice sites. The decreased motion of the LBs decreases the thermal current. Thus, the thermal conductivity $\kappa$ is decreased as we have observed in the range from $C=0.1$ to $C=0.3$ in Fig.\ref{fig3}(a).\\

On the other side, the coefficients $B$ in Eq.(\ref{avgequB}) are proportional to $c^2\tau\sim C\tau$. The enhancement of the anharmonic coefficient $C$, together with the increase of $\tau$, increases the coefficients $B$ in Eq.(\ref{avgequB}). According to the term $\overline{\Delta(\gamma_p^*\xi_j)}/\tau$ in Eq.(\ref{avgequ}), the increase of the coefficients $B$ on the right hand side of Eq.(\ref{avgequ}) increases $\Delta(\gamma_p^*\xi_j)$, which means more LBs move from the $j$-th lattice site to the $p$-th site and take part in the transport with a larger coefficient $C$. This mechanism contributes to the increase of $\kappa$ in the range from $C=0.4$ to $C=0.8$ observed in Fig.~\ref{fig3}(a). Therefore, the increase of $C$ has two effects on the transport of the LBs. One effect is to decrease the thermal current by trapping the LBs on the lattice sites. The other effect is to increase the thermal current by aiding more LBs in number to move in the lattice. The competition between the two effects leads to dependence of $\kappa$ on $C$ in an non-monotonic way.\\

By using the data in Fig.~\ref{fig3}(a), we plot the relation between $\kappa$ and the length $L$ of the system in Fig.~\ref{fig3}(b). For clarity, we show only four plots for the various values of $C$ in Fig.~\ref{fig3}(b). It is clearly shown that the thermal conductivity $\kappa$ starts from a small value and then increases rapidly with the length $L$, showing the finite size effect of $\kappa$. This is because the two ends of the system are correlated through the correlation coefficient $\rho$ introduced in the system. The correlation between the two ends is the basis for the finite size effect of $\kappa$.\\

The length dependence of $\kappa$ comprises two segments. The first segment is for the finite size effect of $\kappa$ as we have illustrated in Fig.~\ref{fig3}(b), say the range from $L=10$ to $L=50$. The finite size effect has been confirmed by the experiments~\cite{Chen13}. The systems in this segment have very small numbers of the atoms and are far from the thermodynamic limitation. For these systems, the MD results are not reliable. Comparably, our theory can clearly capture the finite size effect of $\kappa$ in this segment.\\

The second segment of the length dependence of $\kappa$ is that $\kappa$ increases slowly with the length $L$ of the systems. The thermal conductivity $\kappa$ diverges with the length $L$, and the divergence is the well-known power-law for the FPU chain ~\cite{Lepri03}. For this segment, the length of the system normally is very large beyond several hundreds to several thousands, which has exceeds the computational ability of our theory at the present stage. The data from $L=60$ to $L=100$ in Fig.~\ref{fig3}(b) are not enough for the study of the power-law of $\kappa$, which is still under research for our theory. \\

\subsection{Hopping and Diffusing}
\begin{figure}[ht]
\includegraphics[scale=0.8]{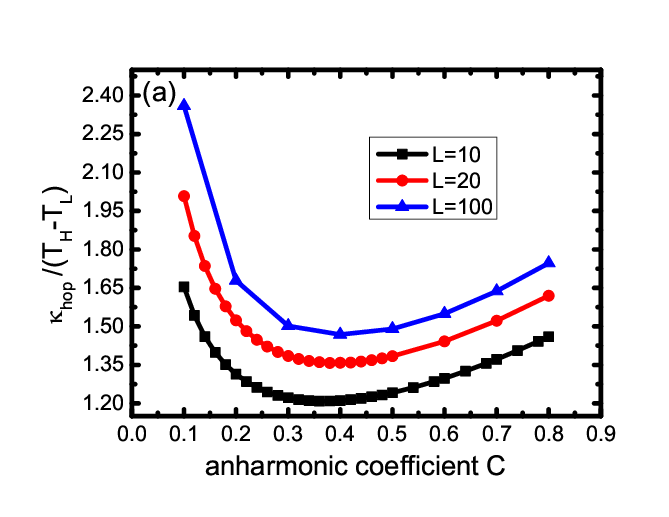}
\includegraphics[scale=0.8]{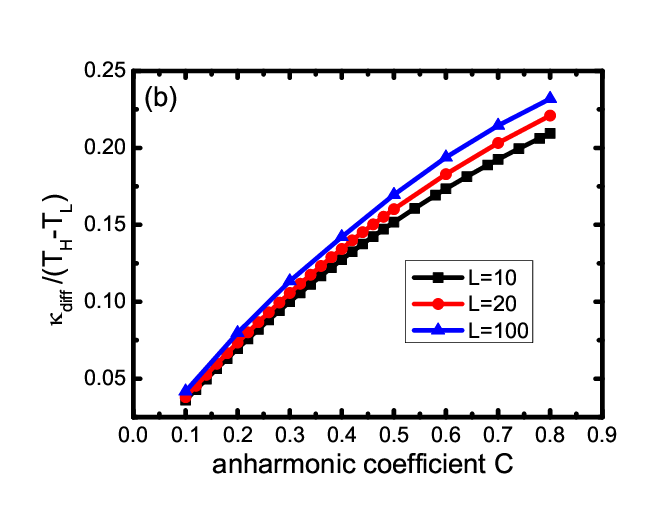}
\caption{Contributions of the hopping and the diffusing of the LBs in the system to the thermal conductivity. (a)The thermal conductivity is dominated by the hopping of the LBs in the system. (b) The thermal conductivity attributed to the diffusing of the LBs in the system increases with the anharmonic coefficient $C$.}
\label{fig4}
\end{figure}

The thermal current defined in Eq.(\ref{Jin}) has three terms. The first two terms are responsible for the hopping of the LBs in the system. The contribution of these two terms to the thermal conductivity is denoted by $\kappa_{hop}$. The third term in Eq.(\ref{Jin}) is for the LBs diffusing in the lattice, which is dependent on the gradient of the LB numbers in the lattice. The contribution of the third term to the thermal conductivity is denoted by $\kappa_{diff}$. We plot the thermal conductivities $\kappa_{hop}$ and $\kappa_{diff}$ as functions of the anharmonic coefficient $C$ in Fig.~\ref{fig4} by varying the length $L$. 
For clarity, we choose only three plots for the various lengths. Fig.~\ref{fig4}(a) is for $\kappa_{hop}$ while Fig.~\ref{fig4}(b) is for $\kappa_{diff}$. It is indicated in Fig.~\ref{fig4} that $\kappa_{hop}$ dominates the thermal conductivity $\kappa$ and the contribution of $\kappa_{diff}$ is negligible.\\

It is observed in Fig.~\ref{fig4}(b) that $\kappa_{diff}$ increases with the anharmonic coefficient $C$. We have revealed in Sec.(\ref{taurho}) that the anharmonic interaction weakens the correlation between the atoms, which decreases the hopping of the LBs and enhances the diffusing, as shown in Fig.~\ref{fig4}(b). It is also observed in Fig.~\ref{fig4}(b) that $\kappa_{diff}$ increases with the length $L$ for a fixed value of $C$. This result suggests that the power-law divergence of $\kappa$ may be attributed to the diffusion of the LBs in the lattice instead of the quantum hopping of the LBs. Due to the limitation of the computational ability, we have not verified this suggestion, which will be our future research topic.

\section{conclusions}
In this study, we introduce local bosons(LBs) as carriers for the thermal transport in a FPU chain. The anharmonic interaction is transformed to a random field, which avoids the enormous boson space for the numerical calculation. By using the open quantum theory, we study the thermal transport in the FPU-$\beta$ lattice. Results show that the anharmonic interaction in the system has two effects on the thermal transport. One effect is to decrease the thermal current by trapping the LBs on the lattice sites. The other effect is to increase the thermal current by enhancing the amount of the LBs for the transport. The competition between the two effects influences the thermal conductivity. The finite size effect of the thermal conductivity has been clearly captured by our theory.\\

The definition of the LBs in our theory is independent of the size and the anharmonic interaction of the system. Comparably, the definition for the velocity of the phonons is vague if the anharmonic interaction is introduced in the Hamiltonian. The MD simulations are not feasible for the finite size effect of the thermal transport in the system, and can not reflect the quantum behaviors of the transport. In this study, our theory can overcome the difficulties occurring to the phonon theory and the MD simulations, and reveal the influence of the anharmonic interaction on the thermal transport.\\

At the present stage, our theory can not be applied to the systems with a large size. Therefore, the power-law divergence of the thermal conductivity in the FPU chain is a challenge to our theory. We will extend the density functional theory for the LBs in large systems as our future work.\\ 

\begin{acknowledgments}
The author kindly acknowledges Ning-Hua Tong and Wenjun Liu from Renmin University of China for discussions .
\end{acknowledgments}

\end{document}